\documentclass[10pt]{revtex4}
\usepackage{amssymb}
\usepackage{latexsym}
\usepackage{epsfig}

\usepackage{amsmath}
\begin{document}
\title{Ghost dark energy in the DGP braneworld}
\author{M. Abdollahi Zadeh$^{1}$\footnote{
m.abdollahizadeh@shirazu.ac.ir} and A.
Sheykhi$^{1,2}$\footnote{asheykhi@shirazu.ac.ir}}
\address{$^1$ Physics Department and Biruni Observatory, College of
Sciences, Shiraz University, Shiraz 71454, Iran\\
$^2$ Research Institute for Astronomy and Astrophysics of Maragha
(RIAAM), P.O. Box 55134-441, Maragha, Iran}

\begin{abstract}
We investigate the ghost model of dark energy in the framework of
DGP braneworld. We explore the cosmological consequences of this
model by determining the equation of state parameter, $\omega_D$,
the deceleration and the density parameters. We also examine the
stability of this model by studying the squared of the sound speed
in the presence/absence of interaction term between dark energy
and dark matter.  We find out that in the absence of interaction
between two dark sectors of the Universe we have
$\omega_D\rightarrow -1$ in the late time, while in the presence
of interaction $\omega_D$ can cross the phantom line $-1$. In both
cases the squared of sound speed $v_s^2$ does not show any signal
of stability. We also determine the statefinder diagnosis of this
model as well as the $\omega_D-{\omega}^{\prime}_{D}$ plane and
compare the results with the $\Lambda$CDM model. We find that
$\omega_D-{\omega}^{\prime}_{D}$ plane meets the freezing region
in the absence of interaction between two dark sectors, while it
meets both the thawing and the freezing regions in the interacting
case.
\end{abstract}
\maketitle

\section{Introduction}
The current acceleration of the Universe expansion which was
strongly confirmed by the type Ia supernova observations
\cite{nova} and also supported by the astrophysical data obtained
from WMAP \cite{WMAP}, indicates the existence of a fluid with
negative pressure, which can overcome the gravity force between
the galaxies and push them to accelerate. It is a general belief
that dark energy (DE) is responsible for such an acceleration,
though its nature and origin is still an open question in the
modern cosmology. There are two approaches for explanation of the
cosmic acceleration. (i) the modified gravity models such as
$f(R)$ gravity \cite{fr} and scalar-tensor theories \cite{BD},
(ii) the idea of the existence of a strange type of energy whose
gravity is repulsive such as the cosmological constant $\Lambda$
\cite{landa} and the dynamical DE models \cite{P,Nojiri20006}.
Against the cosmological constant $\Lambda$ which has constant
equation of state (EoS) parameter $\omega_D=-1$, the further
observations detect a small variation in the EoS parameter of DE
in favor of a dynamical DE with $\omega_D>-1$ in the past and even
$\omega_D<-1$ in the late time \cite{Feng}.

An interesting model for probing the dynamical DE model is the
ghost dark energy (GDE) model proposed in \cite{Ghost}. The
advantages of this model is that it does not introduce any new
degree of freedom in contrast to most DE models that explain the
accelerated expansion by introduction new degree(s) of freedom or
by modifying the underlying theory of gravity. This is important
because, with introducing new degrees of freedom, one needs to
investigate the nature and new consequences in the universe so it
seems to be impressive and economic if we can explain DE puzzle by
using currently known fluids and fields of nature. Actually, GDE
model which is based on the Veneziano ghost in Quantum
Chromodynamics (QCD) can act as the source of DE \cite{Ghost1} and
its existence  are required for resolution of the U(1) problem in
QCD \cite{Witten}. Indeed, the ghosts are decoupled from the
physical states and make no contribution in flat Minkowski space,
but it produces a small vacuum energy density in a dynamic
background or a curved spacetime proportional to $\Lambda
^3_{QCD}H$, where $H$ is the Hubble parameter and $\Lambda_{QCD}$
is QCD mass scale of order a $100 MeV$ \cite{Ohta}. Different
features of GDE have been studied in ample details \cite{Sheykhi}.

Independent of the DE puzzle, for explanation of the cosmic
acceleration, special attention is also paid to extra dimensional
theories, in which our Universe is realized as a $3$-brane
embedded in a higher dimensional spacetime. Based on the
braneworld model, all the particle fields in the standard model
are confined to a four-dimensional brane, while gravity is free to
propagate in all dimension. One of the original model of
braneworld is introduced by Dvali-Gabadadze-Porrati (DGP)
\cite{DGP}, which describes our Universe as a 4D brane embedded in
a 5D Minkowskian bulk with infinite size. In this model the
recovery of the usual gravitational laws on the brane is obtained
by adding an Einstein-Hilbert term to the action of the brane
computed with the brane intrinsic curvature. It is a well known
that the DGP model has two branches of solutions. The
self-accelerating branch of DGP model can explain the late time
cosmic speed-up without recourse to DE or other components of
energy \cite{Def1,Def2}. However, the self-accelerating DGP branch
has ghost instabilities and it cannot realize phantom divide
crossing by itself. To realize phantom divide crossing it is
necessary to add at least a component of energy on the brane. On
the other hand, the normal DGP branch cannot explain acceleration
but it has the potential to realize a phantom-like phase by
dynamical screening on the brane. Adding a DE component to the
normal branch solution brings new facilities to explain late time
acceleration and also better matching with observations. These are
the motivations to add DE to this braneworld setup
\cite{Nozari,brane}. In this work we would like to investigate the
GDE model in the framework of the DGP braneword. This study is of
great importance, since we can incorporate and disclose the
effects of the extra dimension on the evolution of the
cosmological parameters on the brane when the DE source is in the
form of GDE.

This paper is organized as follow. In section \ref{GDE.DGP} we
formulate the GDE model in the context of the DGP braneworld. We
also consider both interacting and noninteracting cases and
explore various cosmological parameters as well as cosmological
planes. Besides the discussion of instability analysis, we study
the $\omega_D-{\omega}^{\prime}_{D}$ plane and properties of
statefinder parameters. We finish with closing remarks in section
\ref{conclusion}.

\section{the GDE in the DGP model}\label{GDE.DGP}
In the DGP cosmology, a homogeneous, spatially flat and isotropic
$3$-dimensional brane which is embedded in a $5$-dimensional
Minkowskian bulk, can be described by the following Friedmann
equation \cite{Wu}
\begin{eqnarray}\label{Fri}
H^2&=&\left(\sqrt{\frac{\rho_m+\rho_D}{3m_p^2}+\frac{1}{4r_c^2}}+\frac{\epsilon}{2r_c}\right)^2,
\end{eqnarray}
or equivalently
\begin{equation}\label{Fri1}
H^2-\frac{\epsilon}{r_c}H=\frac{1}{3m_p^2}(\rho_m+\rho_D),
\end{equation}
where $H={\dot a}/{a}$ is the Hubble parameter, $r_c={m_{\rm
pl}^2}/({2m_{5}^3})$ \cite{Koyama} is the crossover length scale
reflecting the competition between 4D and 5D effects of gravity
and $\epsilon=\pm1$ corresponds to the two branches of solutions
of the DGP model. Before going any further, it is worthy to note
that if $H^{-1}\ll r_c$ (early times) the 4D general relativity is
recovered, otherwise the 5D effect becomes significant. Also
$\epsilon=+1$ corresponds to the self-accelerating solution where
the universe may accelerate in the late time purely due to
modification of gravity \cite{Def1,Def2}, while $\epsilon=-1$ can
produce the acceleration only if a DE component is included on the
brane. Here, to accommodate GDE into the formalism we take
$\epsilon=-1$.

The fractional energy density parameters are defined as
\begin{equation}\label{dimlesspar}
\Omega_{m}=\frac{\rho_{m}}{3m_p^2H^2},
\quad\Omega_{D}=\frac{\rho_{D}}{3m_p^2H^2},\quad \Omega_{r_c}=\frac{1}{4r_c^2H_0^2},
\end{equation}
where $H_0$ is the Hubble parameter at redshift $z=0$. The
Friedmann equation (\ref{Fri1}) can be rewritten in terms
Eq.(\ref{dimlesspar}) as
\begin{eqnarray}\label{Fri2}
\Omega_{m}+\Omega_{D}+2\epsilon\frac{H_0}{H} \sqrt\Omega_{r_c}&=&1.
\end{eqnarray}
We introduce $\Omega_{ DGP} = 2\epsilon
\sqrt{\Omega_{r_c}}{H_0}/{H} $, which comes from the extra
dimension. Thus the Friedmann equation (\ref{Fri2}) can be
reexpressed as
\begin{equation}\label{Fri3}
\Omega_m + \Omega_D +\Omega_{DGP}=1.
\end{equation}
 For the GDE density we have
\begin{equation}\label{DE2}
\rho_D=\alpha H,
\end{equation}
where $\alpha$ is a constant of order $\Lambda ^3_{QCD} $ and
$\Lambda_{QCD}$ is the QCD mass scale \cite{Ohta2011}. Taking the
time derivative of the energy density $\rho_D$ and using
Eq.(\ref{DE2}) we obtain
\begin{equation}\label{dotDE2}
\dot{\rho}_D=\rho_D \frac{\dot{H}}{H}.
\end{equation}
For the FRW universe filled with DE and DM, with mutual
interaction, the energy-momentum conservation law can be written
as
\begin{eqnarray}\label{conm}
&&\dot{\rho}_m+3H\rho_m=Q,\\
&&\dot{\rho}_D+3H(1+\omega_D)\rho_D=-Q,\label{conD}
\end{eqnarray}
where $Q = 3b^2 H(\rho_D+\rho_m)$ is considered as the interaction
term between DE and DM also $b^2$ is the coupling constant of
interaction $Q$.

We know that (i) our Universe is in a DE dominated phase and (ii)
our Universe that is our habitat is stable. These imply that any
variable DE model should result a stable DE dominated universe. So
it is worth investigating the stability of the GDE in DGP
braneworld against perturbation. The intended indicator for
checking the stability of a proposed DE model is to study the
behavior of the squared sound speed ($v_s^2={dP}/{d\rho}$)
\cite{Peebles}. If $v_s^2<0$ we have the classical instability of
a given perturbation because the perturbation of the background
energy density is an oscillatory function and may grow or decay
with time. When $v_s^2 >0$, we expect a stable universe against
perturbations because the perturbation in the energy density,
propagates in the environment. We continue discussion of stability
in the linear perturbation regime where the perturbed energy
density of the background can be written as
\begin{equation}\label{pert1}
\rho(t,x)=\rho(t)+\delta\rho(t,x),
\end{equation}
where $\rho(t)$ is unperturbed background energy density. For the
energy conservation equation ($\nabla_{\mu}T^{\mu\nu}=0$) which
yields \cite{Peebles}
\begin{equation}\label{pert2}
\delta\ddot{\rho}=v_s^2\nabla^2\delta\rho(t,x),
\end{equation}
we encounter two cases. In the first case where $v_s^2 >0$, we
observe an ordinary wave equation which have a wave solution in
the form $\delta \rho=\delta\rho_0e^{-i\omega t+i\vec{k}.\vec{x}}$
(stable universe). In the second case where $v_s^2<0$, the
frequency of the oscillations becomes pure imaginary and the
density perturbations will grow with time as $\delta
\rho=\delta\rho_0e^{\omega t+i\vec{k}.\vec{x}}$ (unstable
universe). Since $v_s^2 $ plays a crucial role in determining the
stability of DE model, we rewrite it in terms of EoS parameter as
\begin{equation}\label{stable}
{v}^{2}_{s}=\frac{\dot P}{\dot\rho} =\frac{\dot\rho_{D}
w_{D}+\rho_{D}\dot w_D}{\dot\rho_{D}(1+u)+\rho_{D}\dot u},
\end{equation}
where $P=P_D$ is the pressure of DE, $\rho=\rho_{m}+\rho_{D}$ is
the total energy density of DE and DM and
$u={\Omega_m}/{\Omega_D}$ is the energy density ration.

On the other sides, Sahni et al., \cite{Sahni} proposed new
geometrical diagnostic pair parameter $\{r,s\}$, known as
statefinder parameter, for checking the viability of newly
introduced DE models. Unlike the physical variables which depend
on the properties of physical fields describing DE models, the
statefinder pair primarily depends on the scale factor and hence
it depends on the metric of the spacetime. The $ r$ and $s$
parameters are defined as \cite{Sahni}
\begin{equation}\label{statefinder}
r=\frac{\dddot{a}}{aH^3},~~~~~~~~~~~~~~ s=\frac{r-1}{3(q-1/2)},
\end{equation}
where $r$ can rewrite as
\begin{equation}\nonumber
r=1+3\frac{\dot{H}}{H^2}+\frac{\ddot{H}}{H^3}.
\end{equation}
and then
\begin{equation}\label{rr2}
r=2q^2+q-\frac{\dot q}{H}.
\end{equation}
Let us note that in the $\{r,s\}$ plane, $s >0$ corresponds to a
quintessence-like model of DE and $s<0$ corresponds to a
phantom-like model of DE. Also the studies on a flat $\Lambda$CDM
model and matter dominated universe (SCDM) show that for these
models $\{r,s\}=\{1,0\}$ and $\{r,s\}=\{1,1\}$, respectively. In
above equations $q$ is the deceleration parameter which is given
by
\begin{equation}\label{q}
q=-1-\frac{\dot{H}}{H^2}.
\end{equation}
In what follows we discuss the $\omega_D-{\omega}^{\prime}_{D}$
plane which introduced by Caldwell and Linder \cite{Caldwell} for
analyzing the dynamical property of various DE models and
distinguish these models (${\omega}^{\prime}_{D}$ represents the
evolution of $\omega_D$). The models can be categorized into two
different classes: (i) ${\omega}^{\prime}_{D}>0$ and $\omega_D<0$
which present the thawing region. (ii) ${\omega}^{\prime}_{D}<0$
and $\omega_D<0$ which present the freezing region. It should be
noted that the $\Lambda$CDM model corresponds to a fixed point
$\{\omega_D=-1,{\omega}^{\prime}_{D}=0\}$ in the
$\omega_D-{\omega}^{\prime}_{D}$ plane. We shall consider the
noninteracting and interacting cases, separatively.

\subsection{Non interacting case}
We start to obtain the cosmological parameters for GDE in the DGP
braneworld by ignoring the interaction term $(Q=0)$. The
deceleration parameter $q$ can be obtained by taking the time
derivative of Eq.(\ref{Fri1}), which lead to
\begin{equation}\label{dotH}
\frac{\dot
H}{H^2}=\frac{-3(1-\Omega_{DGP})+3\Omega_D}{2-\Omega_{DGP}-\Omega_D}.
\end{equation}
Using relation (\ref{q}), we find
\begin{equation}\label{q1}
q=-1-\frac{-3(1-\Omega_{DGP})+3\Omega_D}{2-\Omega_{DGP}-\Omega_D}.
\end{equation}
Inserting Eq. (\ref{dotDE2}) in Eq. (\ref{conD}) we have
\begin{equation}\label{omega}
\omega_D=-1-\frac{1}{3}\frac{\dot H}{H^2},
\end{equation}
where by replacing Eq.(\ref{dotH}) in it, we get
\begin{equation}\label{omega1}
\omega_D=-\frac{1}{2-\Omega_{DGP}-\Omega_D}.
\end{equation}
Also we obtain ${\omega}^{\prime}_{D}$ from the above equation as
\begin{equation}\label{omegaprime}
{\omega}^{\prime}_{D}=-\frac{3(-1+\Omega_{DGP}+\Omega_D)
(\Omega_{DGP}+\Omega_D)}{(-2+\Omega_{DGP}+\Omega_D)^3}.
\end{equation}
Note that in order to find the evolution of  density parameter
$\Omega_D$, we take the time derivative of relation
$\Omega_{D}={\rho_{D}}/{(3m_p^2H^2)}$, after combining the result
with Eqs .(\ref{dotDE2}) and (\ref{dotH}), yields
\begin{equation}\label{Omega}
{\Omega}^{\prime}_{D}=\Omega_D(1+q).
\end{equation}
\begin{figure}[htp]
\begin{center}
\includegraphics[width=8cm]{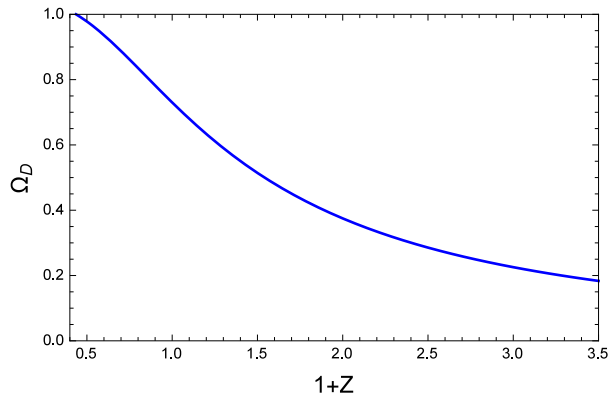}
\caption{The evolution of $\Omega_D$ versus redshift parameter $z$ for
 noninteracting GDE in DGP model . Here, we have taken
$\Omega_D(z=0)=0.73$, $H(z=0)=67$ and $\Omega_{r_c}=0.0003$
}\label{Omega-z1}
\end{center}
\end{figure}
In Fig.~\ref{Omega-z1}, we plot the evolution of $\Omega_D$ versus
redshift parameter $z$. It is obvious that $\Omega_D$ tends to $0$
in the early universe where $1+z\rightarrow\infty$, while at the
late time where $1+z\rightarrow\ 0$, we have $\Omega_D\rightarrow\
1$.
\begin{figure}[htp]
\begin{center}
\includegraphics[width=8cm]{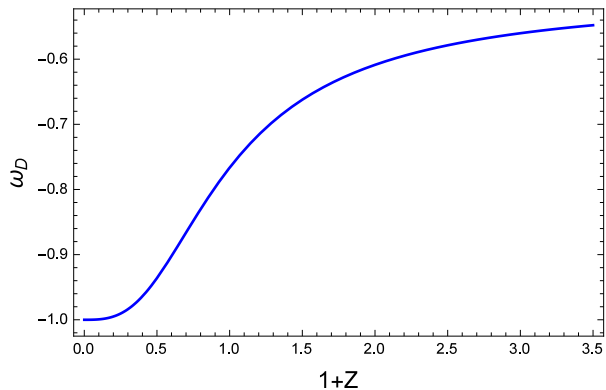}
\caption{The evolution of $\omega_D$ versus redshift parameter $z$ for
 noninteracting GDE in DGP model . Here, we have taken
$\Omega_D(z=0)=0.73$, $H(z=0)=67$ and
$\Omega_{r_c}=0.0003$}\label{w-z1}
\end{center}
\end{figure}
Clearly, Eq.(\ref{omega1}) for the EoS parameter shows that at the
late time where $\Omega_D\rightarrow\ 1$, the EoS parameter mimics
the cosmological constant, namely $\omega_D\rightarrow -1$.
\begin{figure}[htp]
\begin{center}
\includegraphics[width=8cm]{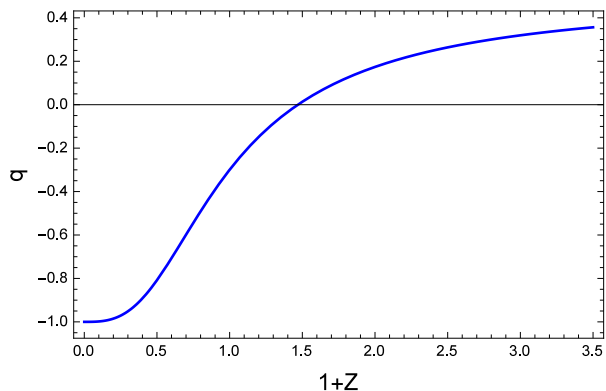}
\caption{The evolution of the deceleration parameter $q$ versus redshift parameter $z$ for
 noninteracting GDE in DGP model . Here, we have taken
$\Omega_D(z=0)=0.73$, $H(z=0)=67$ and
$\Omega_{r_c}=0.0003$}\label{q-z1}
\end{center}
\end{figure}
In Fig.~\ref{q-z1}, the behavior of the deceleration parameter $q$
is plotted and indicates that indeed there is a decelerated
expansion at the early stage of the universe followed by an
accelerated expansion. The energy density ratio is defined as
$u={\Omega_m}/{\Omega_D}$, which by using Eq.(\ref{Fri3}) can be
written
\begin{equation}\label{u}
u=-1+\frac{1}{\Omega_D}(1-\Omega_{DGP}).
\end{equation}
Differentiating Eqs. (\ref{u}) and (\ref{omega1}) and then
substituting the results in Eq. (\ref{stable}) we get the squared
of sound speed as
\begin{equation}\label{stable1}
{v}^{2}_{s}=-\frac{2\Omega_D(-1+\Omega_{DGP}+\Omega_D)}{(-2+\Omega_{DGP})(-2+\Omega_{DGP}+\Omega_D)^2}.
\end{equation}

\begin{figure}[htp]
\begin{center}
\includegraphics[width=8cm]{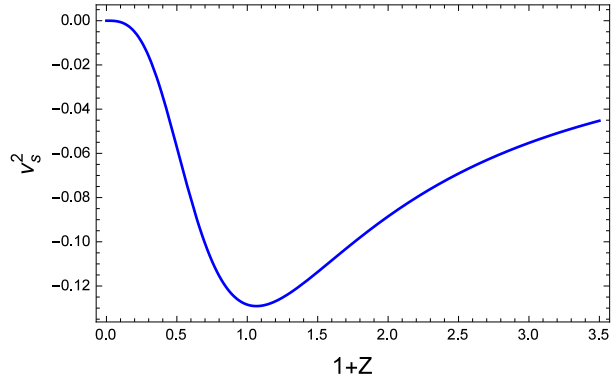}
\caption{The evolution of  the squared of sound speed $v_s^2 $ versus redshift parameter $z$ for
 noninteracting GDE in DGP model. Here, we have taken
$\Omega_D(z=0)=0.73$, $H(z=0)=67$ and
$\Omega_{r_c}=0.0003.$}\label{v-z1}
\end{center}
\end{figure}
The evolution of ${v}^{2}_{s}$ against $z$ for the noninteracting
GDE in the framework of DGP braneworld is plotted in
Fig.~\ref{v-z1}. From graphical analysis of ${v}^{2}_{s}$ one
concludes that this model does not indicate any signal of
stability, that is ${v}^{2}_{s}<0$ during the history of the
universe. We can also find the statefinder parameters $r$ and $s$
by taking derivative of Eq.(\ref{q1}) and using
Eq.(\ref{statefinder}) and Eq.(\ref{rr2}). The results are
\begin{eqnarray}\label{parameterr}
&&r=10+\frac{18}{(-2+\Omega_{DGP}+\Omega_D)^3}+\frac{45}
{(-2+\Omega_{DGP}+\Omega_D)^2}+\frac{36}{-2+\Omega_{DGP}+\Omega_D},\\
&&s=\frac{2(-1+\Omega_{DGP}+\Omega_D)^2}{(-2+\Omega_{DGP}+\Omega_D)^2}.\label{parameters}
\end{eqnarray}
\begin{figure}[htp]
\begin{center}
\includegraphics[width=8cm]{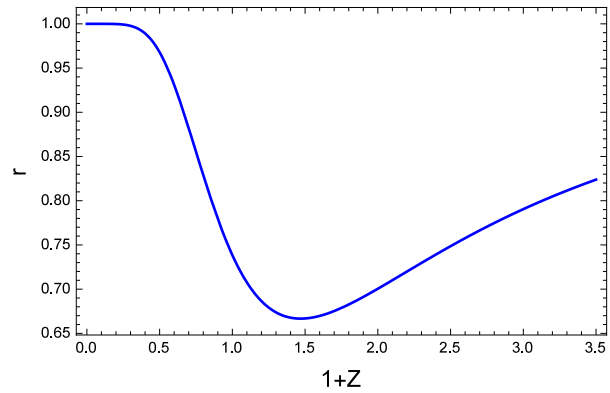}
\caption{The evolution of the statefinder parameter $r$ versus the redshift parameter $z$ for
 noninteracting GDE in DGP model. Here, we have taken
$\Omega_D(z=0)=0.73$, $H(z=0)=67$ and
$\Omega_{r_c}=0.0003.$}\label{r-z1}
\end{center}
\end{figure}
\begin{figure}[htp]
\begin{center}
\includegraphics[width=8cm]{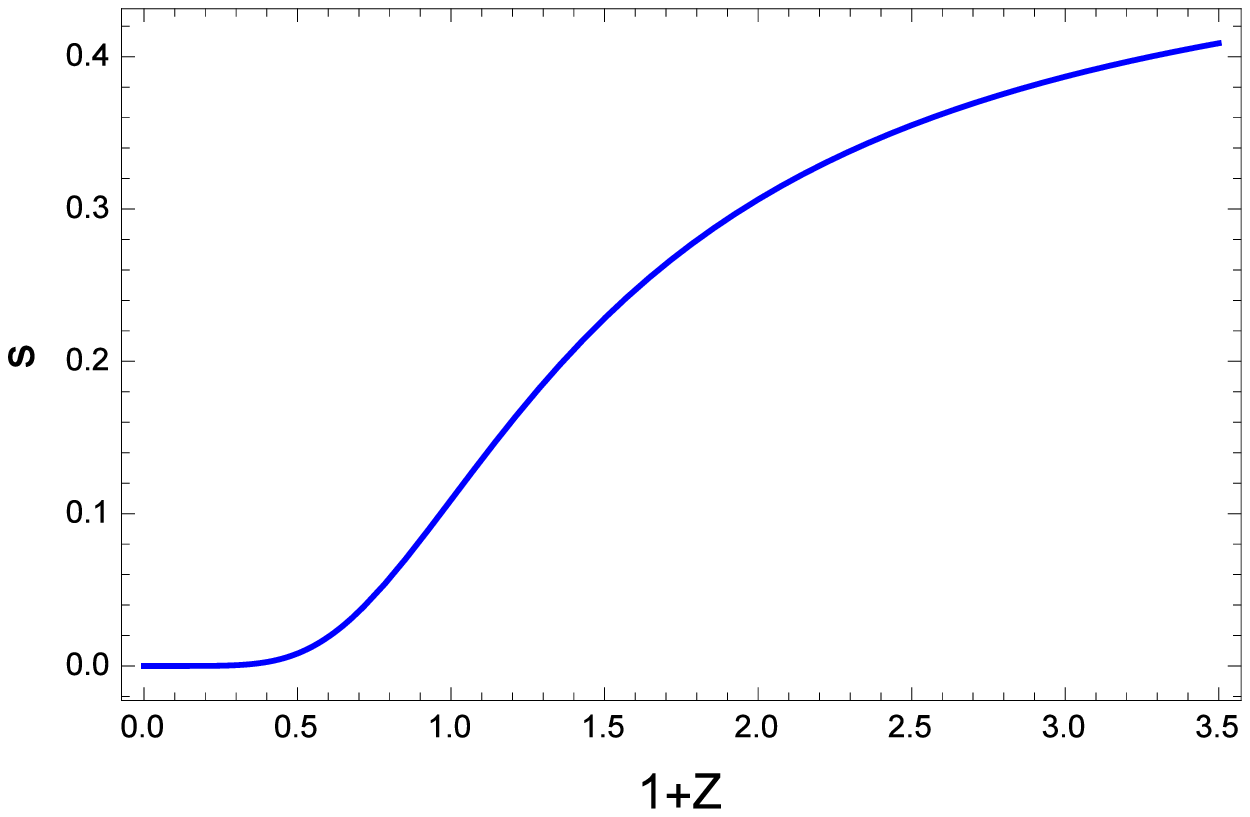}
\caption{The evolution of the statefinder parameter $s$ versus the redshift parameter $z$ for
 noninteracting GDE in DGP model . Here, we have taken
$\Omega_D(z=0)=0.73$, $H(z=0)=67$ and
$\Omega_{r_c}=0.0003$.}\label{s-z1}
\end{center}
\end{figure}
The graphical behavior of the statefinder parameters $\{r,s\}$
given in Eqs. (\ref{parameterr}) and (\ref{parameters}), are
plotted in Figs.~\ref{r-z1} and \ref{s-z1}, showing that at late
time where $\Omega_D\rightarrow\ 1$, we have $\{r,s\}=\{1,0\}$
which implies that GDE mimics the cosmological constant at the
late time, as expected.
\begin{figure}[htp]
\begin{center}
\includegraphics[width=8cm]{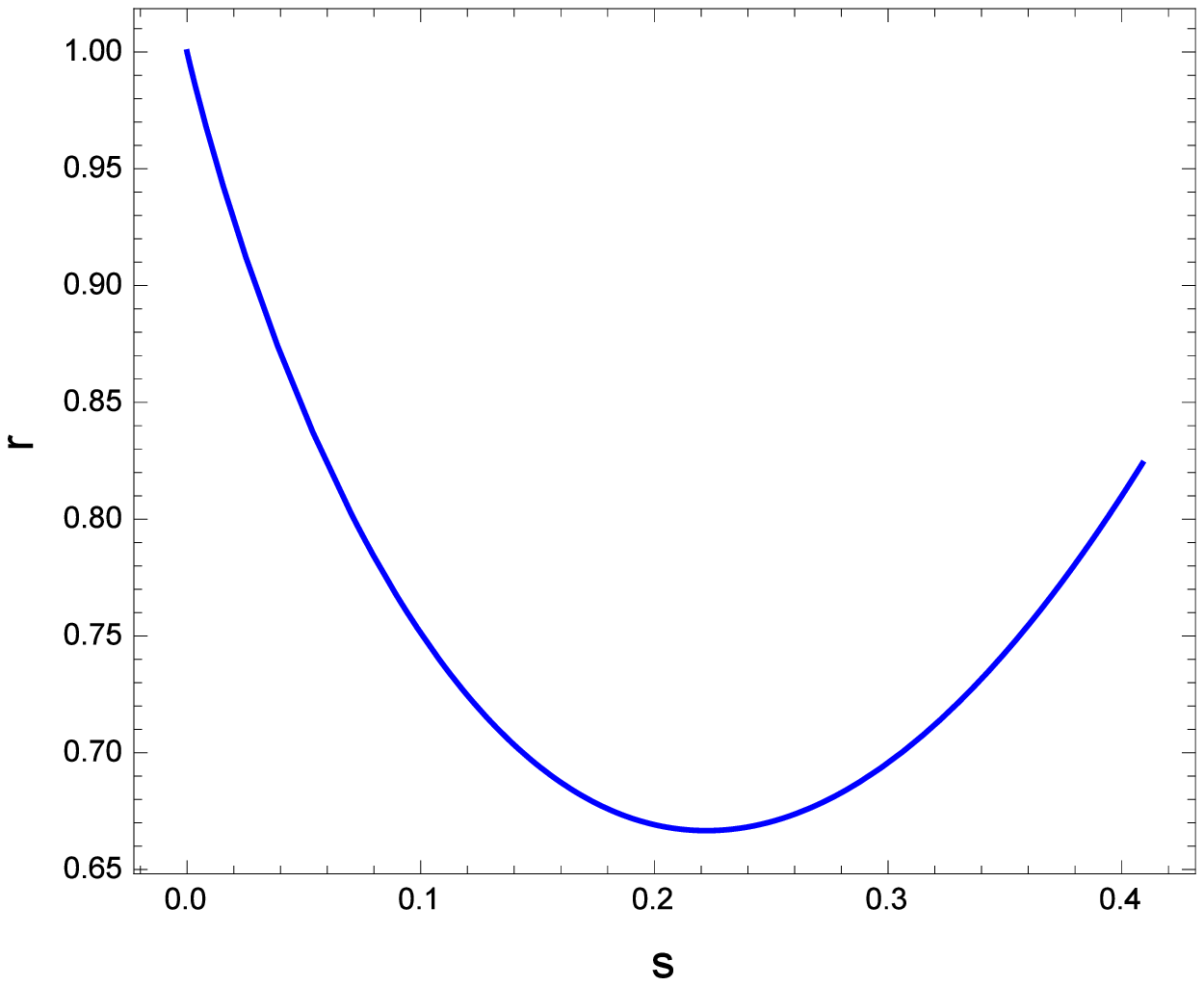}
\caption{The evolution of the statefinder parameter $r$ versus $s$ for
 noninteracting GDE in the DGP model . Here, we have taken
$\Omega_D(z=0)=0.73$, $H(z=0)=67$ and
$\Omega_{r_c}=0.0003.$}\label{rs-z1}
\end{center}
\end{figure}

Let us study the trajectory in the statefinder plane and analyze
this model from the statefinder viewpoint. For this purpose, we
plot the statefinder diagram in the $r-s$ in Fig.~\ref{rs-z1},
which shows the cure gets to the point $\{r,s\}=\{1,0\}$ in the
end, which implies that the model corresponds to the $\Lambda$CDM
model at the late time.
\begin{figure}[htp]
\begin{center}
\includegraphics[width=8cm]{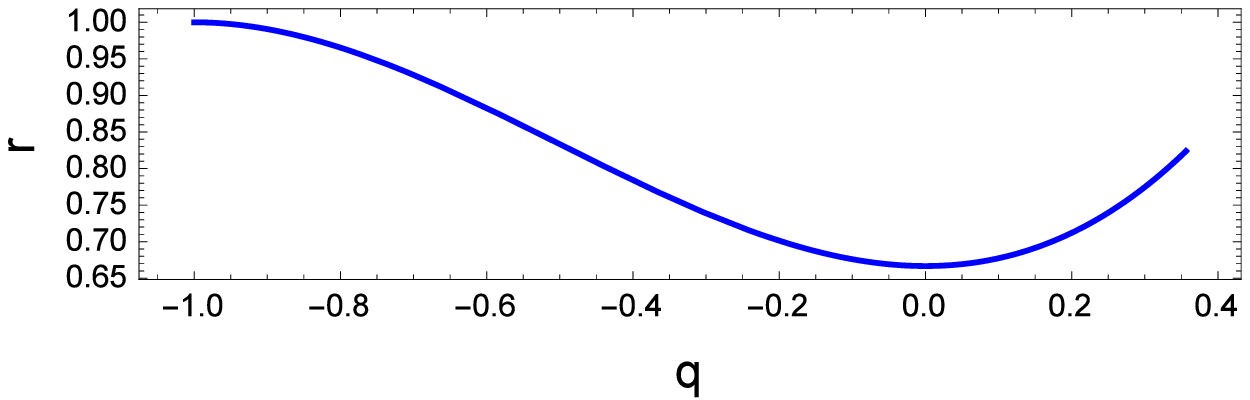}
\caption{The evolution of the statefinder parameter $r$ versus the
deceleration parameter $q$ for noninteracting GDE in the DGP
model . Here, we have taken $\Omega_D(z=0)=0.73$, $H(z=0)=67$ and
$\Omega_{r_c}=0.0003.$}\label{rq-z1}
\end{center}
\end{figure}
For complementarity of the diagnostic, we also plot the
trajectories of statefinder pair $r-q$ in Fig.~\ref{rq-z1} which
ends in the future to $r=1$, $q=-1$ corresponding to the de-Sitter
expansion.
\begin{figure}[htp]
\begin{center}
\includegraphics[width=8cm]{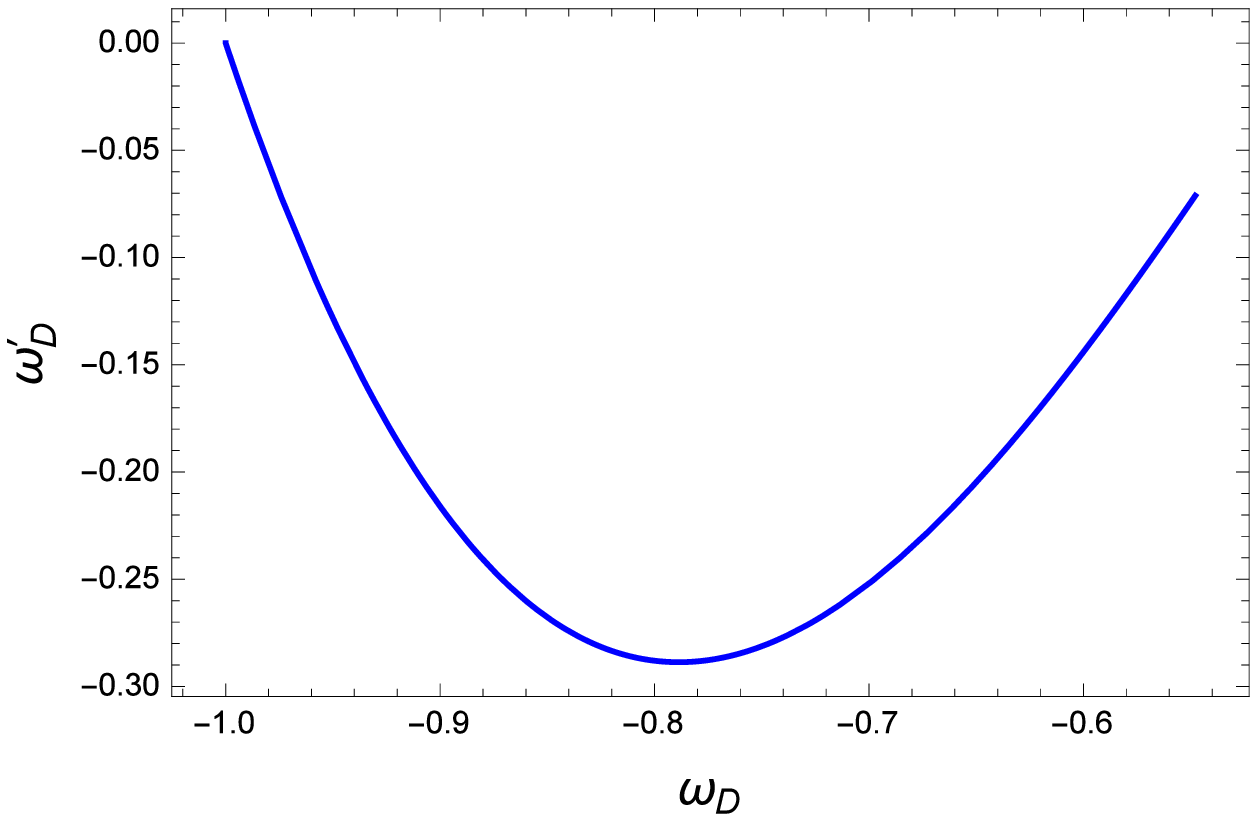}
\caption{The $\omega_D-{\omega}^{\prime}_{D}$ diagram for
noninteracting GDE in the DGP model . Here, we have taken
$\Omega_D(z=0)=0.73$, $H(z=0)=67$ and
$\Omega_{r_c}=0.0003.$}\label{ww-z1}
\end{center}
\end{figure}

The  $\omega_D-{\omega}^{\prime}_{D}$ plane for the noninteracting
GDE in the DGP scenario is shown in Fig.~\ref{ww-z1}. Again, we
see that this plane corresponds to $\Lambda$CDM model, i. e.,
$(\omega_D=-1,{\omega}^{\prime}_{D}=0)$ and the trajectory meets
the freezing region as well.
\subsection{Interacting case}
Differentiating the modified Friedmann equation (\ref{Fri1}) and
using Eqs.(\ref{dotDE2}) and (\ref{conm}) we reach
\begin{eqnarray}\label{dotH1}
\frac{\dot H}{H^2}&=&\frac{3(b^2-1)(1-\Omega_{DGP})+3\Omega_D}{2-\Omega_{DGP}-\Omega_D},\\
q&=&-1-\frac{3(b^2-1)(1-\Omega_{DGP})+3\Omega_D}{2-\Omega_{DGP}-\Omega_D}.
\label{q2}
\end{eqnarray}
Next, the EoS parameter can be determined by substituting
Eq.(\ref{dotDE2}) in the semi-conservation law Eq.(\ref{conD}) and
using Eq.(\ref{dotH1}). We find
\begin{equation}\label{omega2}
\omega_D=\frac{b^2(\Omega_{DGP}-2)(\Omega_{DGP}-1)+\Omega_D}{\Omega_D(-2+\Omega_{DGP}+\Omega_D)}.
\end{equation}
Taking differentiation with respect to $x=\ln a$ from above
equation we get
\begin{equation}\label{omegaprime1}
{\omega}^{\prime}_{D}=-\frac{3[(-1+b^2)(-1+\Omega_{DGP})-\Omega_D]
[b^2(-2+\Omega_{DGP})^2+\left(-\Omega_{DGP}+b^2(-4+3\Omega_{DGP})-\Omega_D\right)\Omega_D]}{\Omega_D(-2+\Omega_{DGP}+\Omega_D)^3},
\end{equation}
where the prime indicates derivative with respect to $x=\ln a$. We
can obtain the equation of motion for $\Omega_D$ as
\begin{equation}\label{Omega1}
{\Omega}^{\prime}_{D}=\frac{3\Omega_D\left(-1+\Omega_D+\Omega_{DGP}+b^2(1-\Omega_{DGP})\right)}{-2+\Omega_{DGP}+\Omega_D},
\end{equation}
\begin{figure}[htp]
\begin{center}
\includegraphics[width=8cm]{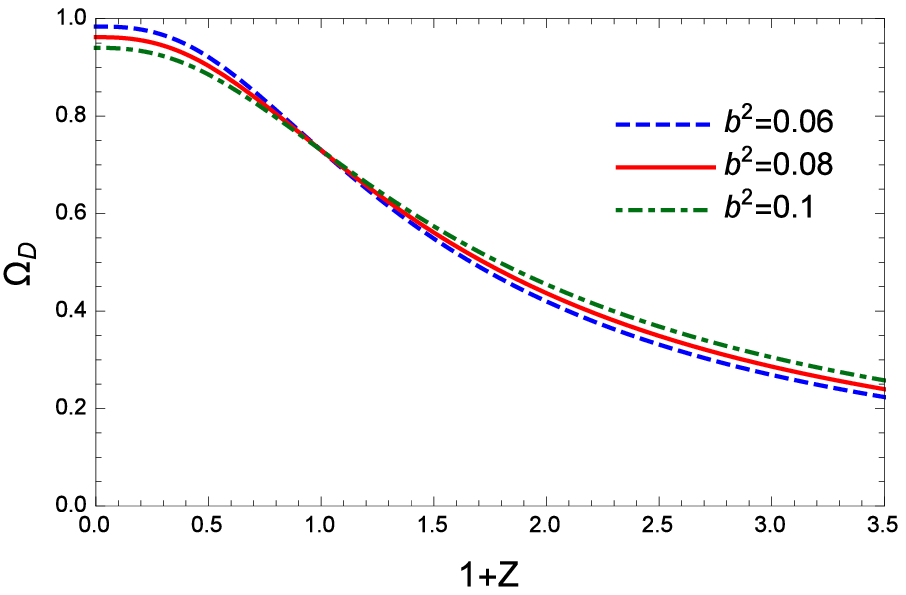}
\caption{The evolution of $\Omega_D$ versus redshift parameter $z$ for
 interacting GDE in the DGP model. Here, we have taken
$\Omega_D(z=0)=0.73$, $H(z=0)=67$ and $\Omega_{r_c}=0.0003$
}\label{Omega-z2}
\end{center}
\end{figure}
To illustrate the cosmological consequences of the interacting GDE
in the DGP braneworld, we plot their evolution in terms of
redshift parameter $z$. In Fig.~\ref{Omega-z2}, we present the
graphical of $\Omega_D$ versus $z$ for the different values of the
coupling constant $b^2$. As expected, we see both
$\Omega_D\rightarrow\ 1$ and $\Omega_D\rightarrow\ 0$ for late
time and early time, respectively.
\begin{figure}[htp]
\begin{center}
\includegraphics[width=8cm]{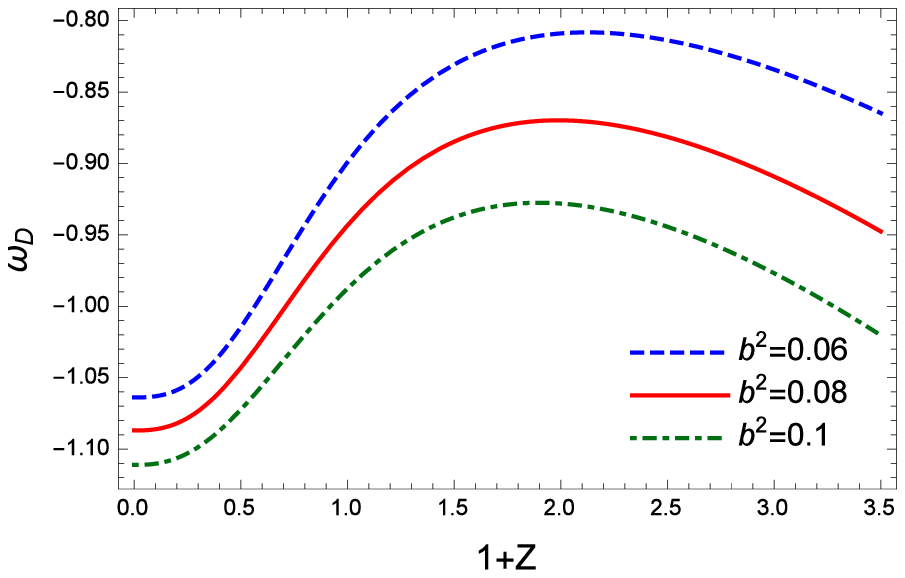}
\caption{The evolution of $\omega_D$ versus redshift parameter $z$ for
 interacting GDE in the DGP model . Here, we have taken
$\Omega_D(z=0)=0.73$, $H(z=0)=67$ and
$\Omega_{r_c}=.0003.$}\label{w-z2}
\end{center}
\end{figure}
The graphical behavior of the EoS parameter for the different
values of $b^2$ shows crossing of phantom line as plotted in
Fig.~\ref{w-z2}. The stability of interacting GDE in DGP model can
obtain by differentiating  with respect time of  Eqs.(\ref{u}) and
(\ref{omega2})
\begin{equation}\label{stable2}
{v}^{2}_{s}=-\frac{b^2[(-2+\Omega_{DGP})^3+
\Omega_D\left(6+(-6+\Omega_{DGP})\Omega_{DGP})\right]+\Omega_D(-2+2\Omega_{DGP}+
2\Omega_D)}{(-2+\Omega_{DGP})(-2+\Omega_{DGP}+\Omega_D)^2}.
\end{equation}
\begin{figure}[htp]
\begin{center}
\includegraphics[width=8cm]{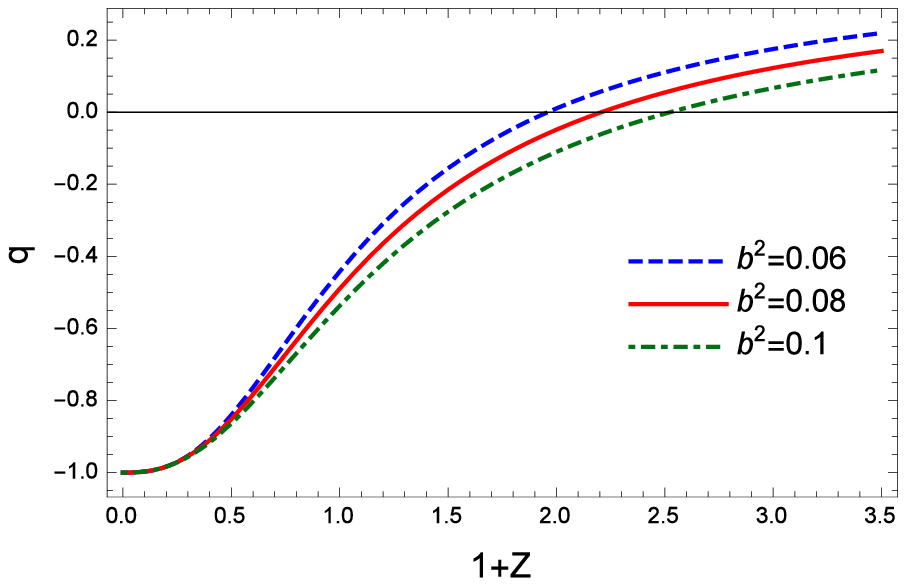}
\caption{The evolution of the deceleration parameter $q$ versus
redshift parameter $z$ for interacting GDE in the DGP model .
Here, we have taken $\Omega_D(z=0)=0.73$, $H(z=0)=67$ and
$\Omega_{r_c}=0.0003.$}\label{q-z2}
\end{center}
\end{figure}

\begin{figure}[htp]
\begin{center}
\includegraphics[width=8cm]{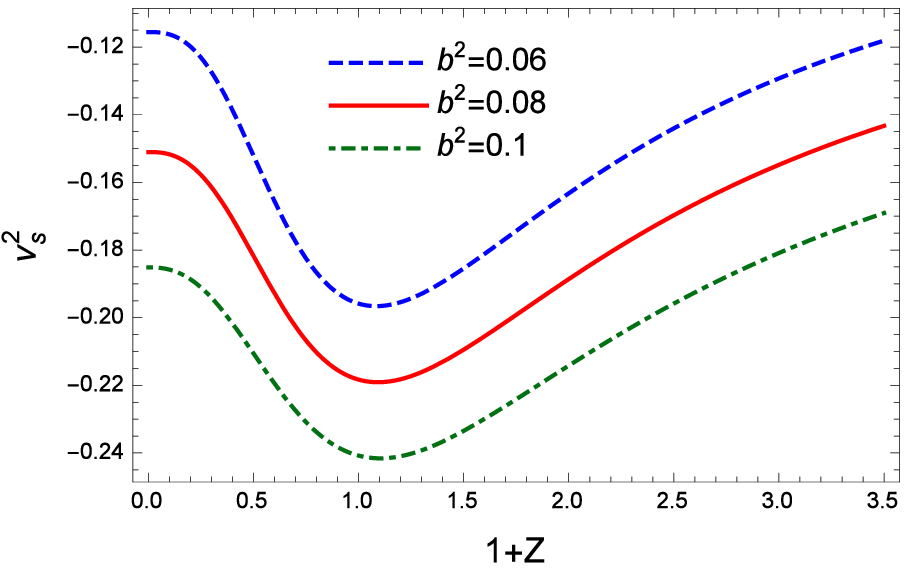}
\caption{The evolution of  the squared of sound speed $v_s^2 $
versus redshift parameter $z$ for interacting GDE in the DGP model
. Here, we have taken $\Omega_D(z=0)=0.73$, $H(z=0)=67$ and
$\Omega_{r_c}=0.0003$}\label{v-z2}
\end{center}
\end{figure}
The evolution of the deceleration parameter $q$ and the squared of
sound speed  $v_s^2 $ versus redshift parameter $z$ are plotted in
Figs.~\ref{q-z2} and \ref{v-z2} respectively. In Fig.~\ref{q-z2},
we see for different values of $b^2$ with the interacting GDE in
DGP model, our universe has a phase transition from deceleration
to an acceleration, while by keeping the same situation in
Fig.~\ref{v-z2}, this universe cannot be stable.  As the value of
$b^2$ decreases the severity of instability also decreases. Like
previous section, the statefinder parameters obtain by taking
derivative of Eq.(\ref{q2}) and using Eq.(\ref{statefinder}) and
Eq.(\ref{rr2})
\begin{equation}\label{parameterr1}
r=10+\frac{18(1+b^2-b^2\Omega_{DGP})}{(-2+\Omega_{DGP}+\Omega_D)^3}+
\frac{9\left(-1+b^2(-1+\Omega_{DGP})\right)\left(-5+b^2(-3+2\Omega_{DGP})\right)}{(-2+\Omega_{DGP}+\Omega_D)^2}+
\frac{9\left(4+b^2(4-3\Omega_{DGP})\right)}{-2+\Omega_{DGP}+\Omega_D},
\end{equation}

\begin{equation}\label{parameters1}
s=2+\frac{2-2b^2(-1+\Omega_{DGP})}{(-2+\Omega_{DGP}+\Omega_D)^2}+
\frac{4+b^2(3-2\Omega_{DGP})}{-2+\Omega_{DGP}+\Omega_D}+\frac{b^2}{2b^2(1-\Omega_{DGp})+\Omega_{DGP}+\Omega_D}.
\end{equation}

\begin{figure}[htp]
\begin{center}
\includegraphics[width=8cm]{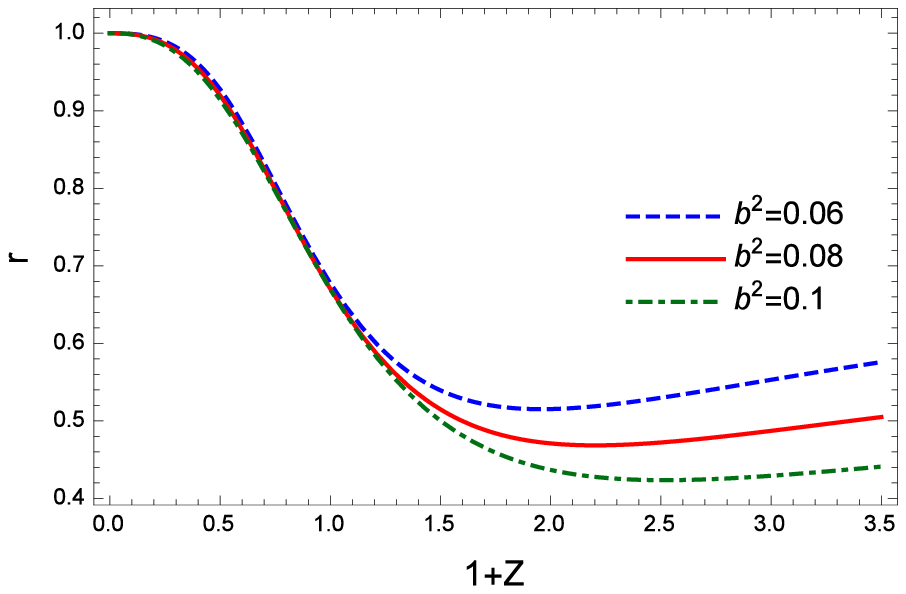}
\caption{The evolution of the statefinder parameter $r$ versus the
redshift parameter $z$ for interacting GDE in DGP model . Here, we
have taken $\Omega_D(z=0)=0.73$, $H(z=0)=67$ and
$\Omega_{r_c}=.0003.$}\label{r-z2}
\end{center}
\end{figure}

\begin{figure}[htp]
\begin{center}
\includegraphics[width=8cm]{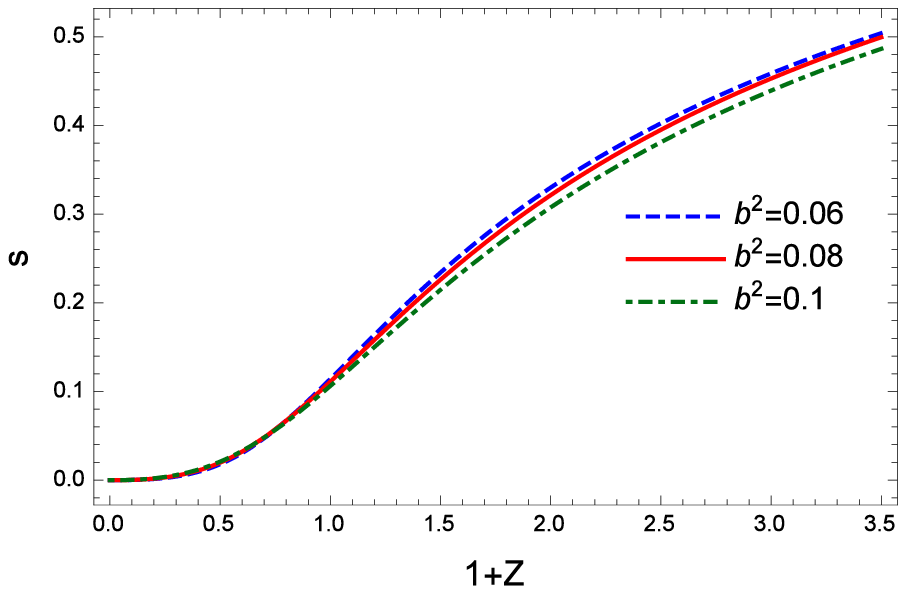}
\caption{The evolution of the statefinder parameter $s$ versus the
redshift parameter $z$ for interacting GDE in the DGP model .
Here, we have taken $\Omega_D(z=0)=0.73$, $H(z=0)=67$ and
$\Omega_{r_c}=.0003.$}\label{s-z2}
\end{center}
\end{figure}
We obtain $\{r,s\}=\{1,0\}$ for $\Lambda$CDM model from
Eqs.(\ref{parameterr1}) and (\ref{parameters1}) in the limiting
case where $b^2=0$, $\Omega_{DGP}=0$ and $\Omega_D\rightarrow\ 1$
(in the late time). Also, Figs.~\ref{r-z2} and \ref{s-z2} show
that $r$ and $s$ are positive through the entire life of the
Universe and turn to 1 and 0 at the late time, respectively.
\begin{figure}[htp]
\begin{center}
\includegraphics[width=8cm]{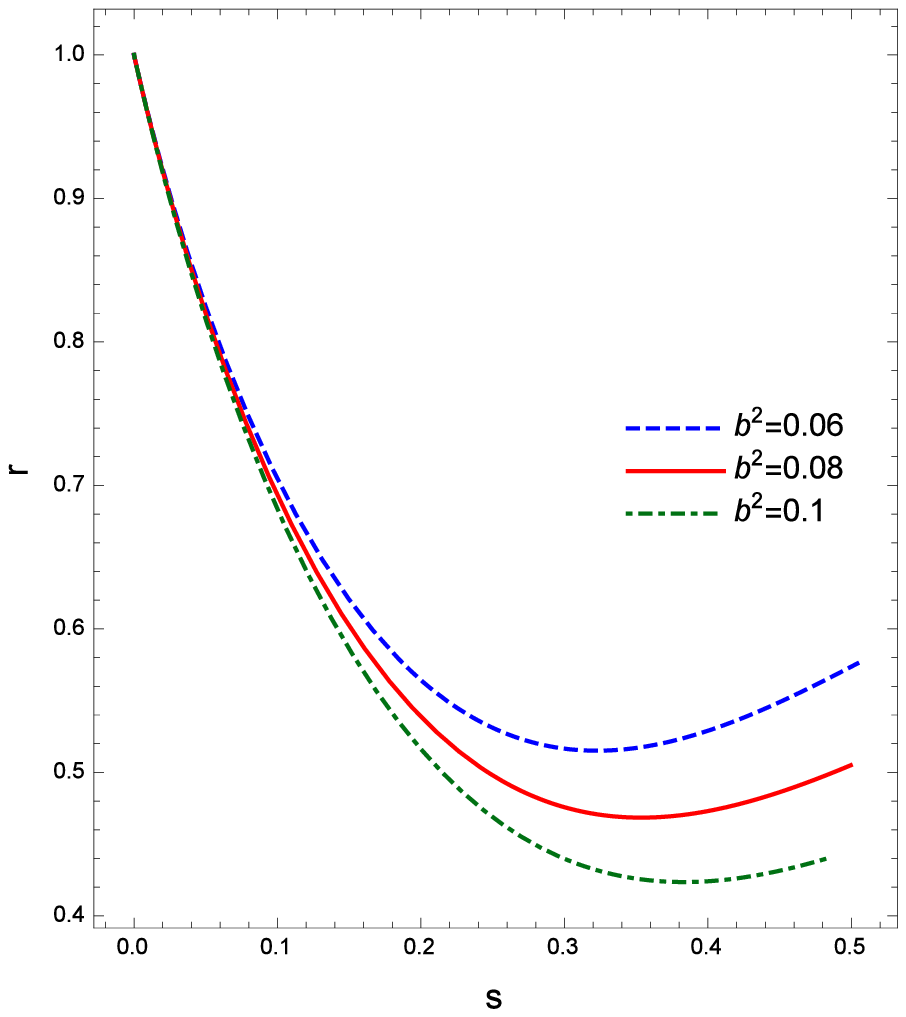}
\caption{The evolution of the statefinder parameter $r$ versus $s$
for interacting GDE in the DGP model. Here, we have taken
$\Omega_D(z=0)=0.73$, $H(z=0)=67$ and
$\Omega_{r_c}=.0003.$}\label{rs-z2}
\end{center}
\end{figure}
The $\{r,s\}$ evolutionary trajectories for the interacting GDE in
the framework of the DGP braneworld for different values of $b^2$
are shown in Fig.~\ref{rs-z2}. From Fig.~\ref{rs-z2}, we can see
that at the late time all curves tend to the $\Lambda$CDM fixed
point $\{r=1, s=0\}$, also different $b^2$, results in different
evolution trajectories of statefinder which states $r$ is smaller
when $b^2$ is larger.
\begin{figure}[htp]
\begin{center}
\includegraphics[width=8cm]{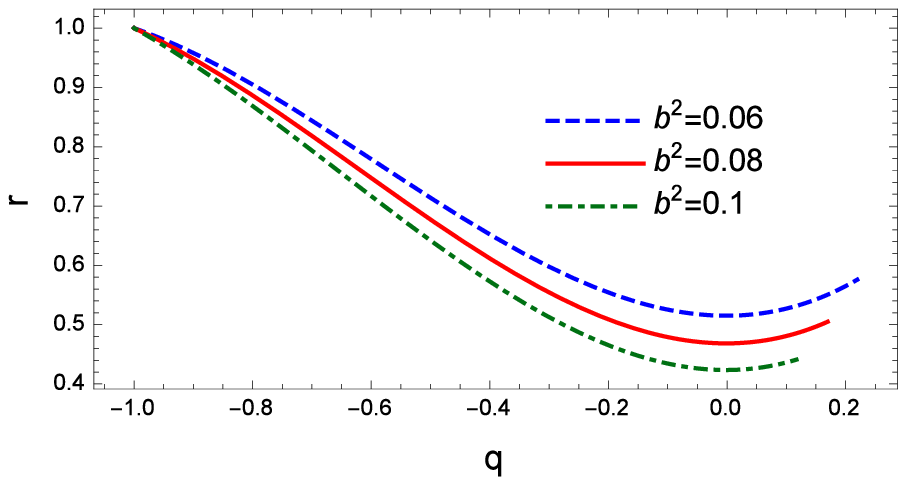}
\caption{The evolution of the statefinder parameter $r$ versus the
deceleration parameter $q$ for interacting GDE in the DGP model .
Here, we have taken $\Omega_D(z=0)=0.73$, $H(z=0)=67$ and
$\Omega_{r_c}=.0003.$}\label{rq-z2}
\end{center}
\end{figure}
The $r- q$ diagrams are plotted for different values of $b^2$ in
Fig.~\ref{rq-z2} which mimics the de Sitter expansion, namely
$r=1$, $q=-1$ in the far future where $z\rightarrow\ 0$.
\begin{figure}[htp]
\begin{center}
\includegraphics[width=8cm]{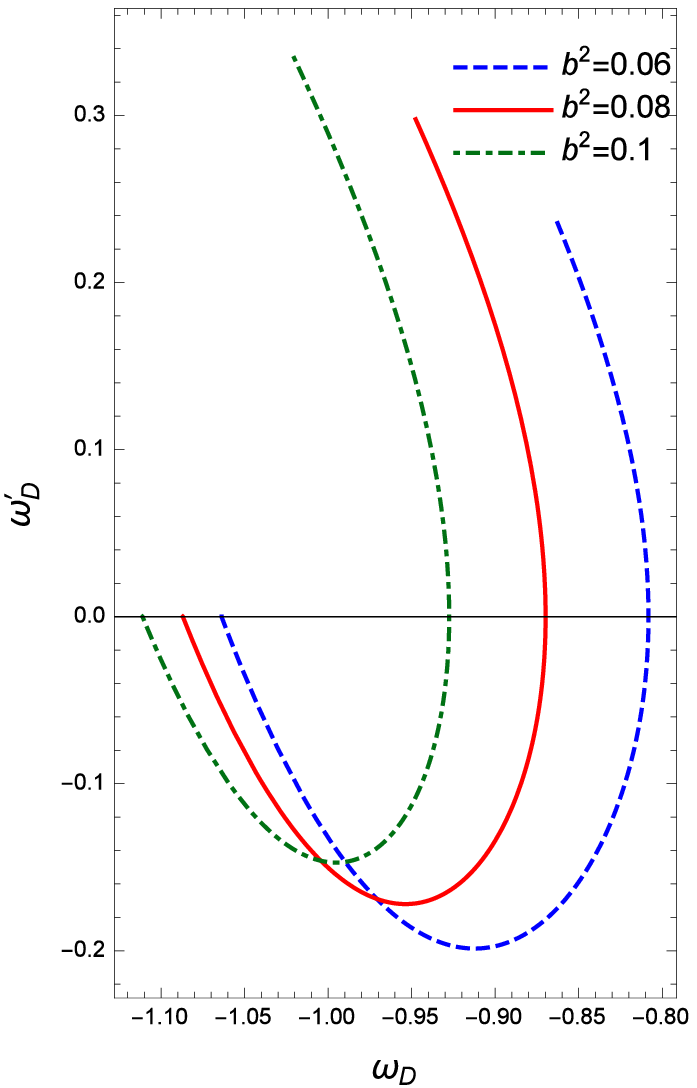}
\caption{The $\omega_D-{\omega}^{\prime}_{D}$ diagram for
interacting GDE in the DGP model. Here, we have taken
$\Omega_D(z=0)=0.73$, $H(z=0)=67$ and
$\Omega_{r_c}=.0003.$}\label{ww-z2}
\end{center}
\end{figure}
In Fig.~\ref{ww-z2}, we plot the $\omega_D-{\omega}^{\prime}_{D}$
plane for different values of $b^2$ which show the trajectories
meet both the thawing and the freezing regions as well.
\section{Closing remarks}\label{conclusion}
We have made a versatile study on both noninteracting and
interacting GDE in the framework DGP model through well-known
cosmological parameters as well as planes. We summarize our
results as follows. For noninteracting case, we have found that
the density parameter tends to zero at the early universe while at
the late time we have $\Omega_D\rightarrow\ 1$. Meanwhile the EoS
parameter cannot cross the phantom line and mimics the
cosmological constant at the late time (Fig.~\ref{w-z1}). We have
shown in that our Universe has a phase transition from
deceleration to an acceleration, though we do not receive any
signal of stability. The statefinder plane shows the trajectory
corresponds to quintessence model ($s>0$ and $r<1$) while at late
time we have $\{r,s\}=\{1,0\}$ for $\Lambda$CDM model as expected.
The $\omega_D-{\omega}^{\prime}_{D}$ plane in Fig.~\ref{ww-z1}
meets the freezing region as well.

For interacting case, we find that the density and the
deceleration parameters as well as the EoS parameter are
consistent with observational data. We have seem that as the value
of $b^2$ decreases the severity of instability decreases. From $r-
s$ plane, we can see that at the late time all cures tend to the
$\Lambda$CDM fixed point $\{r=1, s=0\}$. Besides, for different
values of $b^2$, the different evolution trajectories of
statefinder are shown which indicates that $r$ is smaller when
$b^2$ is larger. The $r-q$ plane is plotted in Fig.~\ref{rq-z2}
which mimics the de Sitter expansion, namely $r=1$, $q=-1$ in the
far future where $z\rightarrow\ 0$. In the end, the
$\omega_D-{\omega}^{\prime}_{D}$ plane exhibits both freezing and
thawing regions of the universe for all values of $b^2$. Again, in
this case $v_s^2<0$ which implies that interacting GDE in the DGP
braneworld is not stable against perturbation.
\acknowledgments{We thank Shiraz University Research Council. This
work has been supported financially by Research Institute for
Astronomy \& Astrophysics of Maragha (RIAAM), Iran.}


\begin{thebibliography}{99}

\bibitem{nova} S. Perlmutter et al., Astrophys. J. {\bf517}, 565 (1999);\\
 P.M. Garnavich et al., Astrophys. J. {\bf493}, L53 (1998);\\
 A.G. Riess et al., Astron. J. {\bf116}, 1009 (1998);\\
 D.N. Spergel et al., Astrophys. J. Suppl. {\bf148}, 175 (2003);\\
 A.G. Riess, Astrophys. J. {\bf607}, 665 (2004);\\
 P.J. Peebles, B. Ratra, Rev. Mod. Phys. {\bf75}, 559 (2003).
\bibitem{WMAP} G.F. Hinshaw et al. ApJS {\bf208}, 19
(2013).

\bibitem{fr} S. Capozziello, V. F. Cardone, S. Carloni and A. Troisi, Int.
J. Mod. Phys. D {\bf12}, 1969 (2003);\\
S. M. Carroll, V. Duvvuri, M. Trodden and M. S. Turner,
Phys. Rev. D {\bf70}, 043528 (2004);\\
S. Nojiri, S.D. Odintsov, Int. J. Geom. Meth. Mod. Phys.{\bf 4}
, 115 (2007);\\
M. Sadegh Movahed, S. Baghram and S. Rahvar, Phys. Rev.
D {\bf76}, 044008 (2007);\\
S. Baghram, M. Sadegh Movahed and S. Rahvar, Phys. Rev.
D {\bf80}, 064003 (2009);\\T.P. Sotiriou and V. Faraoni Rev. Mod. Phys. {\bf82}, 451 (2010);\\
S. Nojiri and S.D. Odintsov, Phys. Rep. {\bf505}, 59 (2011);\\ M.
Tegmark et al., Astrophys. J. {\bf606}, 702 (2004);\\ M. Kowalski
et al., Astrophys. J. {\bf686}, 749 (2008);\\ E. Komatsu et al.,
Astrophys. J. Suppl. Ser. {\bf180}, 330 (2009);\\ P. A. R. Ade et
al., (Planck Collaboration), Astron. Astroph. {\bf594}, A13
(2016);\\G. Papagiannopoulos, S. Basilakos, J. D. Barrow, A.
Paliathanasis, Phys. Rev. D {\bf97}, 024026 (2018)

\bibitem{BD} L. Amendola, Phys. Rev. D {\bf60}, 043501 (1999);\\
J. P. Uzan, Phys. Rev. D {\bf59}, 123510 (1999);\\
T. Chiba, Phys. Rev. D {\bf60}, 083508 (1999);\\
N. Bartolo andM. Pietroni, Phys. Rev. D {\bf61}, 023518
(2000);\\V. Faraoni, \textit{Cosmology in Scalar-Tensor Gravity},
Kluwer, Boston, (2004); \\ E. Elizalde, S. Nojiri, S. D. Odintsov,
P. Wang, Phys. Rev. D {\bf71}, 103504 (2005);  \\S. Nojiri, S. D.
Odintsov, Gen. Relativ. Gravit. {\bf38}, 1285 (2006) ;\\  R.
Gannouji, et al., JCAP {\bf0609}, 016 (2006);\\N. Banerjee, D.
Pavon, Phys. Lett. B {\bf647}, 447 (2007);\\A. Sheykhi,  Phys.
Lett. B {\bf681}, 205 (2009);\\ A. Sheykhi,
 Phys. Rev. D {\bf81}, 023525 (2010);\\
A. Sheykhi, M. Jamil,  Phys. Lett. B {\bf694}, 284
(2011);\\ E. Ebrahimi, A. Sheykhi,  Phys. Lett. B {\bf706}, 19
(2011);\\ A. Sheykhi, E. Ebrahimi, and Y. Yousefi,
 Can.
J. Phys. {\bf91}, 662  (2013);\\K. Karami, A. Sheykhi, M. Jamil,
Z. Azarmi, M. M. Soltanzadeh, Gen. Relativ. Grav.{\bf43},27
(2011);\\ A.
Pasqua, S. Chattopadhyay, Astrophys. Space Sci. {\bf348}, 284
(2013);\\ V. Fayaz, Astrophys. Space Sci. {\bf361}, 86 (2016)\\ P.
Kumar, C.P. Singh,  Astrophys. Space Sci. {\bf362}, 52
(2017);\\Singh, C.P.  Kumar, P. Int J Theor Phys {\bf56}, 3297
(2017) ;\\F. Felegary, F. Darabi, M. R. Setare, Int. J. Mod. Phys.
D. {\bf27}, 1850017 (2018).

\bibitem{landa} V. Sahni and A. Starobinsky, Int. J. Mod. Phy. D {\bf9}, 373
(2000).

\bibitem{P} M. Li, Phys. Lett. B {\bf603}, 1 (2004);\\
S. D. H. Hsu, Phys. Lett. B {\bf594}, 13 (2004);\\
D. Pavon, W. Zimdahl, Phys. Lett. B {\bf628}, 206 (2005) ;\\  B. Guberina, R. Horvat, H. Stefancic, JCAP
{\bf 0505}, 001  (2005);\\ B. Guberina, R. Horvat, H. Nikolic,
Phys. Lett. B {\bf 636},80 (2006);\\ H. Li, Z. K. Guo, Y. Z.
Zhang, Int. J. Mod. Phys. D {\bf15}, 869 (2006) ;
\\ Q. G. Huang, Y. Gong, JCAP {\bf0408}, 006 (2004) ;\\
J. P. B. Almeida, J. G. Pereira, Phys. Lett. B {\bf636}, 75 (2006)
;
\\  Y. Gong, Phys. Rev. D {\bf70}, 064029 (2004) ; \\ B. Wang, E.
Abdalla, R. K. Su, Phys. Lett. B {\bf611}, 21 (2005) ;\\
 M. R. Setare, S. Shafei, JCAP {\bf09}, 011 (2006) ;\\
M. R. Setare, Eur. Phys. J. C {\bf50}, 991 (2007) ;\\ M. R.
Setare, JCAP
{\bf0701},023 (2007) ;\\ M. R. Setare, Phys. Lett. B {\bf654}, 1 (2007);\\
 M. R. Setare, E. C. Vagenas, Phys. Lett. B {\bf666}, 111 (2008) ;\\
 M. R. Setare, E. N. Saridakis,
Phys. Lett. B {\bf671}, 331 (2009) ;\\ C. Wetterich, Nucl. Phys. B
{\bf 302}, 668 (1988);\\ B. Ratra and J. Peebles, Phys. Rev. D
{\bf 37}, 3406 (1988);\\ R. R. Caldwell, Phys. Lett. B {\bf 545}23
(2002);\\ R. R. Caldwell, M. Kamionkowski and N. N. Weinberg,
Phys. Rev. Lett. {\bf 91}, 071301 (2003);\\ Shin'ichi Nojiri,
Sergei D. Odintsov, Phys.Lett. B {\bf 562}, 147 (2003);\\ T.
Chiba, T.
Okabe, and M. Yamaguchi, Phys. Rev. D {\bf 62}, 023511 (2000);\\
C. Armendariz-Picon, V. F. Mukhanov and P. J. Steinhardt, Phys.
Rev. Lett. {\bf 85}, 4438 (2000);\\ C. Armendariz-Picon, V. F.
Mukhanov and P. J. Steinhardt, Phys. Rev. D {\bf 63}, 103510
(2001);\\ A. Y. Kamenshchik, U. Moschella and V. Pasquier,  Phys.
Lett. B {\bf 511}265 (2001);\\ M. C. Bento, O. Bertolami, and A.
A. Sen,
Generalized Chaplygin Gas,  Phys. Rev. D {\bf 66}, 043507 (2002);\\
A. Sheykhi,  Class. Quantum Gravit. {\bf 27}, 025007 (2010)

\bibitem{Nojiri20006} Shin'ichi Nojiri, Sergei D. Odintsov,  Gen.Rel.Grav. {\bf 38}, 1285 (2006);\\
R. G. Cai, Phys. Lett. B {\bf 657}, 228 (2007);\\ H. Wei and R. G. Cai ,Phys. Lett. B {\bf 660}, 113 (2008);\\
H. Wei and R. G. Cai, Phys. Lett. B {\bf 663}, 1 (2008);\\ J. P.
Wu,
D. Z. Ma, and Y. Ling, Phys. Lett.B {\bf 663}, 152 (2008);\\ J. Zhang, X. Zhang, and H. Liu, Eur. Phys. J. C {\bf 54}, 303 (2008);\\
 A. Sheykhi, Phys. Lett. B {\bf 680}, 113 (2009);\\ A. Sheykhi, Phys. Lett. B {\bf 682}, 329 (2010);\\
I. Duran, L. Parisi ,Phys. Rev. D {\bf85}, 123538 (2012);\\F. Yu,
Jing-Fei Zhang,  Theor. Phys. {\bf59}, 243 (2013) 243;\\P.
Pankunni, Titus K. MATHEW, Int. J. Mod. Phys. D {\bf23}, 1450024
(2014);\\Y. Hu, M. Li, N. Li, Z. Zhang , JCAP{\bf08}, 012
(2015);\\HL. Li, JF. Zhang, L.Feng et al. Eur. Phys. J. C {\bf77},
907 (2017) ;\\Ze. Zhao, Shuang. Wang, Sci.China Phys.Mech.Astron.
{\bf61},039811
(2018) ;\\A. Al Mamon, Int. J. Mod. Phys. D {\bf26}, 1750136 (2017);\\
 Sh. Wang, Yi. Wang, M. Li, Physics Reports {\bf 696},1 (2017) ;\\
 M. Abdollahi Zadeh, A. Sheykhi, H. Moradpour, Int. J. Mod. Phys. D {\bf26}, 8
 (2017).



\bibitem{Feng} B. Feng, X. L. Wang and X. M. Zhang, Phys. Lett. B {\bf607},
35 (2005);\\ U. Alam, V. Sahni and A. A. Starobinsky, JCAP
{\bf0406}, 008 (2004);\\ D. Huterer and A. Cooray, Phys. Rev. D
{\bf71}, 023506 (2005).


\bibitem{Ghost} F. R. Urban and A. R. Zhitnitsky, Phys. Lett. B {\bf688}, 9 (2010);\\
 Phys. Rev. D {\bf80}, 063001 (2009);\\ JCAP {\bf0909}, 018 (2009);\\ Nucl. Phys. B {\bf835}, 135 (2010).

\bibitem{Ghost1} G. Veneziano, Nucl. Phys. B {\bf159}, 213 (1979).


\bibitem{Witten} E. Witten, Nucl. Phys. B {\bf156}, 269 (1979);\\
C. Rosenzweig, J. Schechter, C.G. Trahern, Phys. Rev. D {\bf21}, 3388 (1980).

\bibitem{Ohta} N. Ohta, Phys. Lett. B {\bf695}, 41 (2011).

\bibitem{Sheykhi} A. Sheykhi, M. Sadegh Movahed, Gen. Relativ.
Gravit. {\bf44}, 449 (2012);\\ A. Sheykhi, A. Bagheri, Europhys.
Lett. {\bf95}, 39001 (2011);\\ E. Ebrahimi, A. Sheykhi, Int. J.
Mod. Phys. D, Vol. {\bf20}, No. 12
2369 (2011);\\ E. Ebrahimi, A.Sheykhi, H. Alavirad, Cent. Eur. J. Phys. Vol {\bf11},  No.7, 949 (2013).

 C-J. Feng, X-Z. Li, P. Xi, JHEP
{\bf1205}, 046 (2012);\\ C. J. Feng, X.-Z. Li, X.-Y. Shen, Phys.
Rev. D {\bf87}, 023006 (2013);\\ C.-J. Feng, X.-Z. Li, X.-Y. Shen,
Mod. Phys. Lett. A {\bf27}, 1250182 (2012);\\ S. Nojiri, S.D.
Odintsov, Phys. Rev. D {\bf72}, 023003 (2005);\\ S. Capozziello,
V.F. Cardone,E. Elizalde,  S. Nojiri, S.D. Odintsov, Phys. Rev. D
{\bf73}, 043512 (2006);
\\M.Z. Khurshudyan, A.N. Makarenko, Astrophys Space Sci {\bf 361} 187.(2016);\\M. Malekjani, T. Naderi,
F. Pace, MNRAS {\bf453}, 4148 (2015);\\E. Ebrahimi, H. Golchin, A.
Mehrabi, S. M. S. Movahed IJMPD, {\bf 26},1750124  (2017);\\ M.
Abdollahi Zadeh, A. Sheykhi, H. Moradpour, Int. J. Theor. Phys.
{\bf 56}, 3477 (2017).


\bibitem{DGP} G.R. Dvali, G. Gabadadze,M. Porrati, Phys.Lett. B {\bf485} 208
(2000).

\bibitem{Def1}  C. Deffayet, Phys. Lett. B \textbf{502}, 199 (2001).
\bibitem{Def2}  C. Deffayet and G. Dvali, Phys. Rev. D {\bf65}, 044023 (2002).

\bibitem{Nozari} K. Nozari, N. Behrouz, A. Sheykhi, Int. J. Theor. Phys. {\bf 52},
2351 (2013).


\bibitem{brane} J. Dutta, S. Chakraborty, M. Ansari, Mod. Phys. Lett. A {\bf25}, 3069 (2010);\\
 D. Jibitesh et al. Int.J.Theor.Phys. {\bf50}, 2383 (2011);\\S. Ghaffari, et al. Phys.Rev. D{\bf91}, 023007 (2015);\\
 H. Farajollahi, et al. Astrophys.Space Sci. {\bf348}, 253 (2013);\\ Sh. Rani, A. Jawad, Int. J. Mod. Phys. D
{\bf25},1650102 (2016);\\ Y. Aguilera, A. Avelino, N. Cruz, S.
Lepe, F. Pena, Eur. Phys. J. C  {\bf74}, 3172 (2014);\\ N. Cruz,
S. Lepe, F. Pena, A. Avelino, Eur. Phys. J. C {\bf72}, 2162
(2012);\\ S. Ghaffari, M. H. Dehghani, A. Sheykhi, Phys. Rev. D
{\bf89}, 123009 (2014);\\ A. Jawad, Astrophys Space Sci  {\bf360},
52 (2015);\\ A. Jawad, et al. Eur. Phys. J. Plus  {\bf131}, 236
(2016);\\ A. Jawad, Ines G. Salako, Eur. Phys. J. Plus {\bf130},
198 (2015).


\bibitem{Wu} X. Wu, R.G. Cai, Z.H. Zhu,  Phys. Rev. D {\bf 77}, 043502
(2008).

\bibitem{Koyama} K. Koyama, Gen. Relativ. Gravit. {\bf40}, 421 (2008);\\
M. Li, X. Li, S. Wang and Y. Wang, Commun. Theor. Phys. {\bf56},
525 (2011) .

\bibitem{Ohta2011} N. Ohta, Phys. Lett. B {\bf695}, 41 (2011).

\bibitem{Peebles} P. J. E. Peebles and B. Ratra, Rev. Mod. Phys. {\bf75}, 559 (2003).

\bibitem{Sahni} V. Sahni, T. D. Saini, A. A. Starobinsky and U. Alam, JETP Lett. {\bf77}, 201 (2003).

\bibitem{Caldwell} R. Caldwell and E. V. Linder, Phys. Rev. Lett. {\bf95},  141301 (2005).

\end{thebibliography}
\end{document}